\documentclass[12pt]{article} \textheight 23cm
\usepackage{graphicx}
\usepackage{amsmath}
\usepackage{color}

\begin{document} \begin{center}
{\large Crack Formation in the Presence of an Electric Field in Droplets of Laponite Gel    }\\ \vskip 0.5cm
Tajkera Khatun$^1$, Tapati Dutta$^2$ and Sujata Tarafdar$^{1*}$\\
\vskip 0.5cm

$^1$ Condensed Matter Physics Research Centre, Physics Department, Jadavpur University, Kolkata 700032, India\\

$^2$ Physics Department, St. Xavier's College, Kolkata 700016, India\\

Tajkera Khatun: tajkerakhatu88@gmail.com\\
Tapati Dutta: tapati$\_$mithu@yahoo.com\\
Sujata Tarafdar$^*$: Corresponding author, sujata$\_$tarafdar@hotmail.com\\ 

\end{center}
\vskip .5cm
\noindent {\bf Abstract}\\ When a colloidal gel dries through evaporation, cracks are usually formed, which often reveal underlying processes at work during desiccation. Desiccating colloid droplets of few hundred $\mu l$ size show interesting effects of pattern formation and cracking which makes this an active subject of current research. Since aqueous gels of clay are known to be strongly affected by an electric field, one may expect crack patterns to exhibit a field effect.  In the present study we allow droplets of laponite gel to dry under a radial electric field. This gives rise to highly reproducible patterns of cracks, which depend on the strength, direction and time of exposure to the electric field. For a continuously applied DC voltage, cracks always appear on dissipation of a certain constant amount of energy. If the field is switched off before cracks appear, 
the observed results are shown to obey a number of empirical scaling relations, which enable us to predict the time of appearance and the number of cracks under specified conditions.
 \vskip .5cm
 
 	
 \section{Introduction}	
 The study of drying droplets is an interesting subject which leads to pattern formation, phase segregation, buckling, cracking and other phenomena\cite{review,buckling,moutushi,anarelli1}. 
 Such studies are particularly useful for biological fluids where they help in medical diagnosis\cite{sobac,brutin}. The shape of the droplet, hence wetting properties of the fluid on the substrate\cite{brutin-langmuir}, the evaporation rate\cite{evaporation}, temperature and humidity all affect the process. Evaporating drops have also been studied under a strong static electric field\cite{colsua14} where changes in shape have been observed. Another area where considerable research is going on is the study of dessication crack patterns on layers of clay or other granular material\cite{lucas,daniels,pauchard1,so,lopes}. Here a host of remarkable phenomena such as memory effects\cite{naka}, magnetic\cite{pauchard} and electric field effects in DC \cite{tajkera} and AC \cite{ac} have been observed. 
 
\begin{figure}[h]
\begin{center}
\includegraphics[width=14.0cm, angle=0]{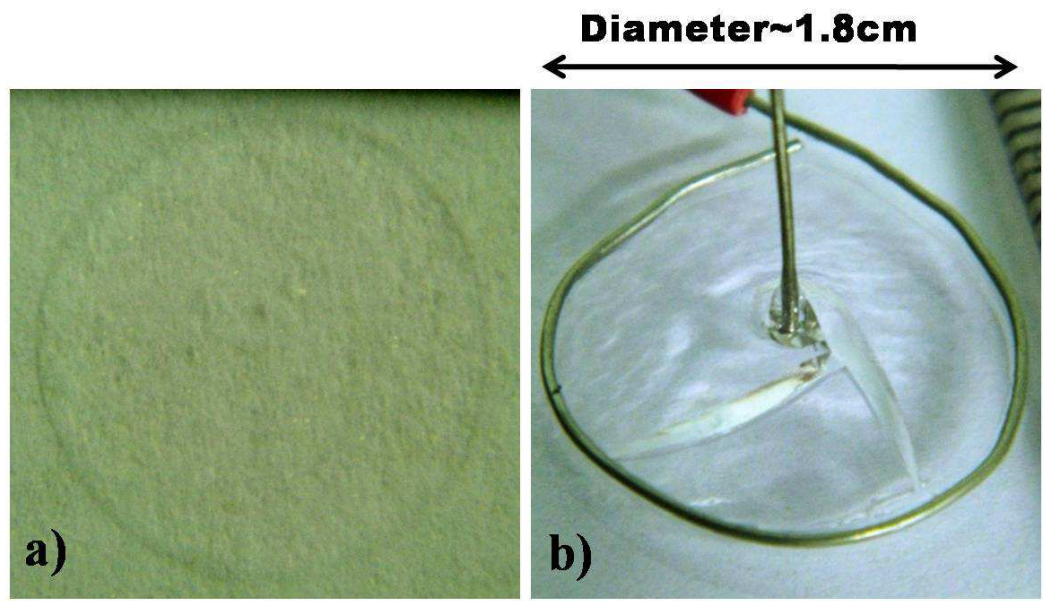}
\end{center}
\caption{a) shows a dried droplet of laponite on an acrylic surface, a uniform film is formed with no cracks. b) shows a drying droplet of laponite with electrodes fitted but no field applied. Cracks appear randomly, without any definite pattern.}
\label{nofield}
\end{figure}

 In the present paper we report experiments on crack formation in droplets of laponite gel under a static electric field. The small size of the droplets creates highly reproducible experimental conditions and the convex geometry of the sessile drop is a feature not present in earlier experiments on larger systems\cite{tajkera,mal1,mal2}. Earlier studies show that droplets of complex fluids often crack on drying \cite{sobac,brutin,yakhno,bardakov,tarasevich}, but we find that a drop of unconfined laponite gel placed on a glass or acrylic surface dries as a uniform film without crack formation (figure \ref{nofield} a)). However, if the droplet is confined within a boundary, cracks appear during desiccation (figure \ref{nofield} b)). In the presence of a radial DC electric field, the pattern and time of appearance of the cracks depend on the direction and strength of the field as well as on the duration of exposure to the field. We find that the cracks always appear at the positive terminal and their formation is affected by the history of  earlier exposure to an electric field even when it has been applied for a very short time. We have studied the formation of crack patterns for four different voltages $V$ (in the range 5 - 12 Volt,) varying duration of application of field $\tau$ (which ranges  from seconds to minutes and finally hours i.e. throughout the experiment). 
 
\begin{figure}[h]
\begin{center}
\includegraphics[width=14.0cm, angle=0]{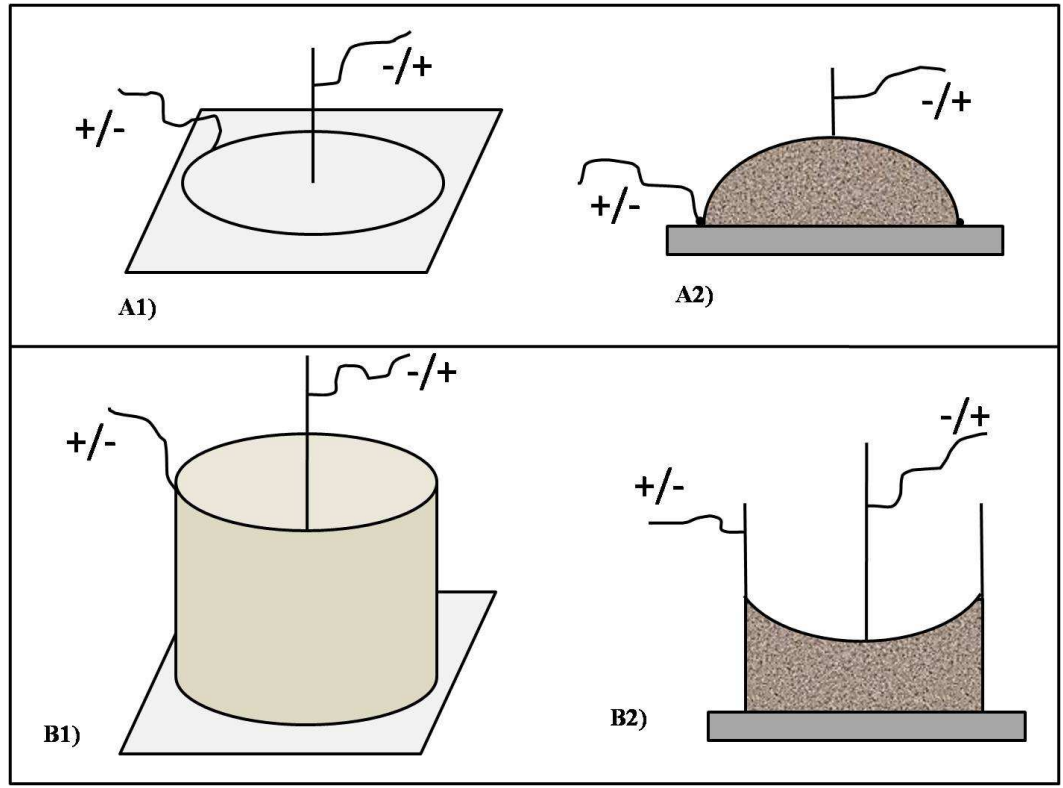}
\end{center}
\caption{A1) and A2) show schematically, the set up and the corresponding profile of drying droplets under a radial field (with CP or CN) respectively. B1) and B2) show schematically, the set up and the corresponding profile of laponite gel under cylindrical geometry (with CP or CN) respectively. }
\label{setup}
\end{figure}

 \begin{figure}[]
\begin{center}
\includegraphics[width=14.0cm, angle=0]{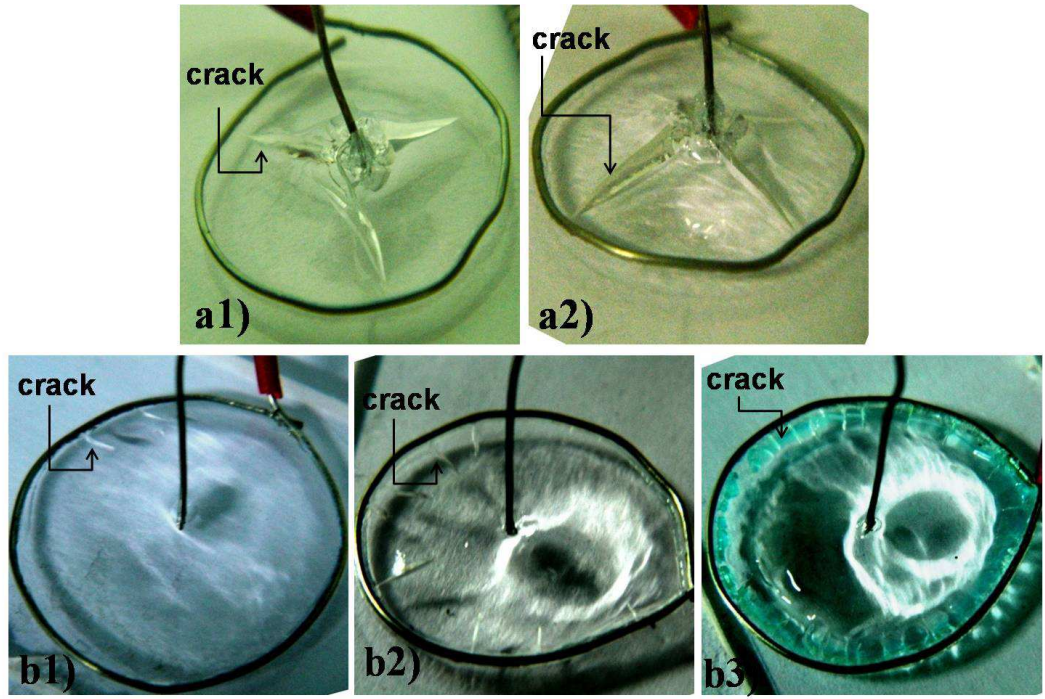}
\end{center}
\caption{Final crack patterns: a1) and  a2) show the crack patterns in drying droplet of laponite with field duration $\tau$ = 1min and 30 sec respectively in CP for 12V. There are almost always 3 cracks formed at the central electrode in CP for all voltages ($V$) and all field duration ($\tau$) as shown in a). In very rare case there are 4 cracks as shown in b). b1), b2) and  b3) show the crack patterns with field duration $\tau$ = 15sec, 1min and 5min respectively in CN for 10V. The number of cracks (indicated by arrows) increases with $\tau$.}
\label{tau-variation-cp-cn}
\end{figure}  
 
 The results are striking.
A number of scaling relations are found, which involve the time of first appearance of cracks ($t_a$) and the total number of final cracks ($n_f$) as function of $V$ and $\tau$. The master curves obtained from the scaling enable us to predict features of crack patterns appearing for any DC voltage applied for any given duration of time. A preliminary version of this study was presented at the workshop - DROPLET-2013 \cite{poster}.

\section{Materials and Methods}
Laponite RD (Rockwood Additives) is used for the present set of experiments. For concentrations above 2 \% it forms a gel when stirred in tap water, but in deionised water higher concentrations are needed for gel formation \cite{net}. An acrylic sheet is used as substrate. During our experiment temperature and humidity varied between 23 $^o$C - 26 $^o$C and 36 \% -56 \% respectively.

0.625g of laponite RD is added to 10 ml of deionised water while it is on a magnetic stirrer. The mixture is stirred for 5-10 sec and then a droplet of diameter $\sim 18 mm$  is placed on an acrylic sheet. The droplet dries as a uniform film, without crack formation (figure(\ref{nofield} a)).

 \begin{figure}[h]
\begin{center}
\includegraphics[width=16.0cm, angle=0]{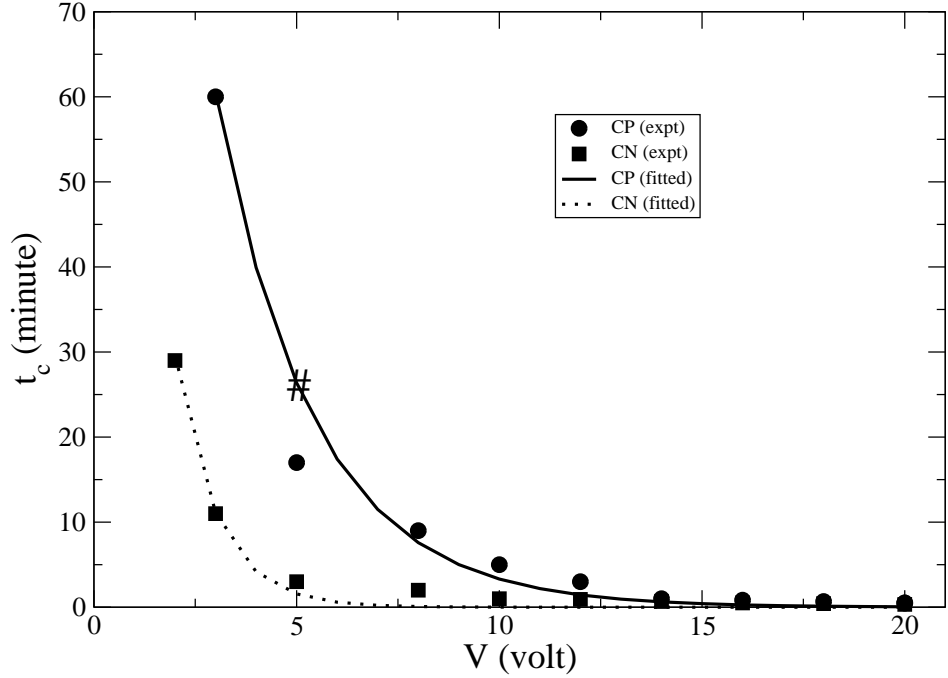}
\end{center}
\caption{Time of appearance of first crack ($t_c$) vs applied voltage ($V$). Symbols are experimental points and solid and broken lines are fitted from equation(\ref{tc}). \# represents the value of $t_c$ for 5V CP calculated from equation(\ref{tc}).}
\label{tc_vs_appliedv}
\end{figure}

To apply DC field to the droplet, a circular ring made of aluminium wire serves as the peripheral electrode (figure \ref{setup} A1), A2)). It is placed on the sheet and the droplet is then deposited inside the ring. Another straight wire of the same material, acting as central electrode is placed at the centre of the droplet. After waiting for 3-4 minutes till the solution spreads out evenly and gels, the power is turned on. The central electrode may be positive (abbreviated as CP) or negative (abbreviated as CN). If the power is not turned on in this set up, cracks appear as the droplet dries, but these have no definite pattern (figure(\ref{nofield} b)).

There is some change in the shape of the laponite drop due to the presence of the ring-like outer electrode. The approximate contact angle of the laponite drop on the acrylic sheet is measured as $\sim 60^o$ while in the presence of electrodes (outer: ring-like and central: rod-like) it is $\sim 74^o$.
   
 \begin{figure}[h]
\begin{center}
\includegraphics[width=16.0cm, angle=0]{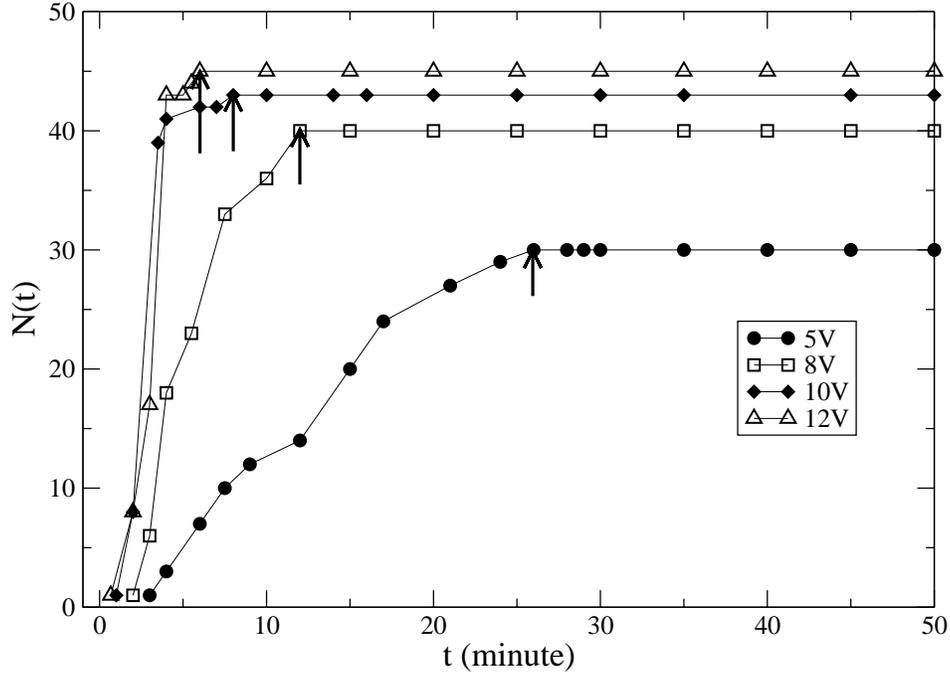}
\end{center}
\caption{$N(t)$ vs. $t$ plots for CN condition for four different voltages($V$). Arrows mark the saturation points ($t_{sat}$, $N_{sat}$).}
\label{Ntsat}
\end{figure}

\subsection{Desiccation under a continuous DC Voltage: Characteristic times - $t_c$ and $t_{sat}$}
A voltage $V$ volts (5V, 8V, 10V, 12V) is applied to the droplet from a DC power supply for 4-5 hours, until there is no further change in the crack pattern. 
There are two easily identifiable  `characteristic times' which depend on the applied voltage. At the critical time 
$t_c$  the first crack  appears. Then the number of cracks $N(t)$ continues to increase subsequently upto a  `saturation time' defined as $t_{sat}$, when $N(t)$ reaches the maximum value $N_{sat}$ for a particular $V$.
$t_c$ and $t_{sat}$  are noted for both CP and CN configurations for different $V$. Variation of the current $I(t)$ across the sample is also noted as a function of time for different $V$. A milli-ammeter is used for this purpose.

\subsection{Crack appearance when field is switched off at $\tau$}
 
 Now we apply the voltage $V$ ($V$ = 5V, 8V, 10V, 12V) to the droplet from a DC power supply  for a duration $\tau$ which is less than $t_c$. We take $\tau$ equal to 15 sec, 30 sec, 1 minute, 3 minutes, 5 minutes and 10 minutes for both CP and CN configurations. As $t_c$'s are different for different magnitudes and directions of the field, the maximum $\tau$ may vary with the magnitude and direction of the field.
 The time of first appearance of cracks $t_a$ is noted for different $V$ and $\tau$.

 The final number of cracks $n_f$ which appear at positive electrode is noted after the field is switched off at time $\tau$ which is allowed to vary form 0 to $t_{sat}$.

 \section{Results}
 When an electric field is applied to a droplet, drying starts and cracks appear first at the positive electrode. Cracks do not in general appear from the negative electrode. Very rarely a few cracks form at the negative electrode, if at all, they appear at a later time. We further note that at the negative electrode water separates out initially, which dries up after some time. These observations are similar to earlier work on larger circular systems \cite{tajkera}.
 
  It is interesting to note that if a laponite-water drop is deposited on an acrylic sheet, no cracks form at all. However, only the presence of electrodes, without any applied field causes cracks to appear randomly. So the adhesion between the sample and a lateral boundary (here, the electrodes) seems a necessary condition for crack formation. The crack pattern in the absence of electric field is shown in figure \ref{nofield}b). 
  
  Figure(\ref{tau-variation-cp-cn}) shows crack patterns for CP and CN conditions. For CP, cracks appear from central electrode, propagate towards the outer ring and almost reach the outer electrode (figure(\ref{tau-variation-cp-cn}) a1, a2). No branches appear. For CN, cracks appearing from the outer ring proceed towards the central electrode, but cannot reach it (figure(\ref{tau-variation-cp-cn}) b1, b2 and b3). One or two cracks may appear at a later time from the central electrode in CN condition. Generally for CN, the central portion of the droplet remains almost free of cracks.

 \subsection{Critical times $t_c$, $t_{sat}$, final number of cracks $n_f$}
 The critical time $t_c$ for crack appearance when the field is always on is plotted as a function of $V$ in figure(\ref{tc_vs_appliedv}) for both CP and CN. It is seen that the curve for CN runs much below CP and both can be fitted by an exponential curve of the form
 
 \begin{equation}
 t_{c} = t_0 exp( -\frac{V}{V_0})
  \label{tc}
 \end{equation}
 
 $t_0$ = 210 minutes is the experimentally observed time of crack appearance in the absence of any electric field. The only free parameter $V_0$ of the equation, may depend on details of the set up and ambient conditions. It ($V_0$) is equal to $0.415V$ for CP, and $0.98V$ for CN conditions. All the experimental points match the fitted data quite well. The only exception is the point corresponding to 5V in CP. There may be some error for this point as we discuss later.

 \begin{figure}[h]
\begin{center}
\includegraphics[width=16.0cm, angle=0]{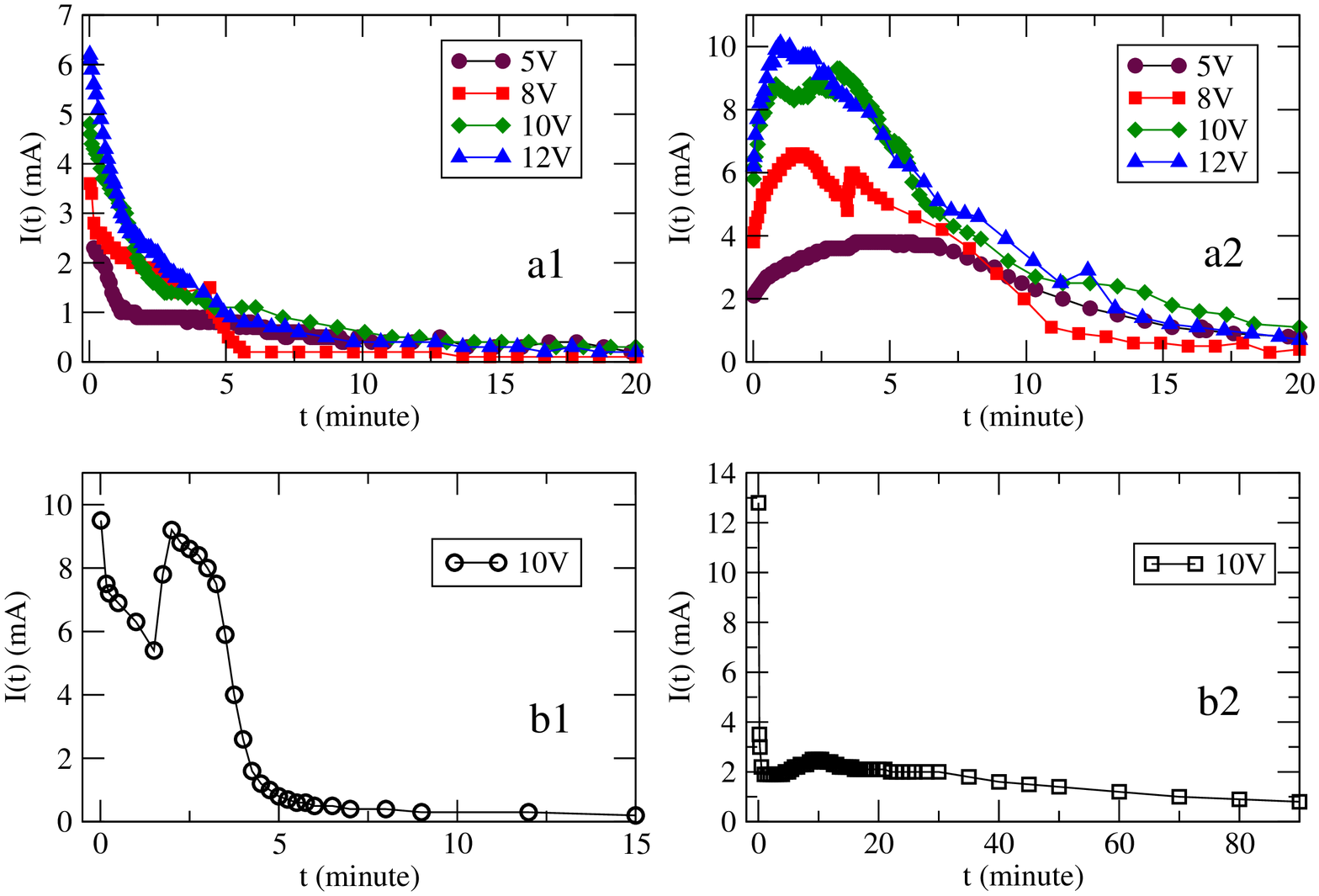}
\end{center}
\caption{Current through the sample I(t) as function of time t. a1) and a2) show results for the droplets (figure \ref{setup} A1, A2) for CP and CN respectively. b1) and b2) show results for the concave meniscus (figure \ref{setup} B1, B2) for CP and CN respectively.}
\label{current_droplet_cp_cn}
\end{figure}

 When a field corresponding to $V$ is on throughout the experiment, cracks start to appear at the  $t_c$ corresponding to that $V$ and their number $N(t)$ increases until a saturation value $N_{sat}$ is reached at time $t = t_{sat}$. $N_{sat}$ increases with $V$ and $t_{sat}$ decreases with $V$ as shown in figure(\ref{Ntsat}).

 Variation of the current $I(t)$ with time is shown for all experiments in figure (\ref{current_droplet_cp_cn} a1, a2). These results show that for CP, the current starts from an initial value depending on $V$ and falls rapidly reaching almost zero in 15-20 min. For CN, though the current starts at nearly the same value as for CP (for a particular $V$), it rises significantly before falling sharply towards zero. We offer an explanation for this peculiar behaviour in the following section and verify it through auxiliary experiments.

\subsection{Auxiliary experiments}
\label{current} 
 We first try to explain the peculiarly different  variations of the current $I(t)$ with time in the case of CP and CN configurations (figure \ref{current_droplet_cp_cn} a1, a2). The initial resistance of the droplet is found to be $\sim$ 2 k$\Omega$. 
It seems quite natural that as the drop hardens with time its resistance should increase and the current should fall continuously, at a constant $V$. 
This is indeed observed in CP, but why does the current increase initially in CN, before starting to fall?

We must remember that the drop is not a homogeneous system before drying and there are regions where excess water (or sol) allows easy paths for ion migration. It is always observed \cite{tajkera} that excess water starts to appear at the negative electrode a few seconds after the DC field is switched on, this water dries up later. For the hemispherical sessile droplet, the position of the central electrode is higher than the peripheral electrode (figure(\ref{setup}) A1, A2). With CP, excess water emitted at the periphery, remains there until it dries up, but with CN, excess water appears at the center and flows towards the periphery. This causes a conducting film of fluid to span the surface of the drop in CN for a short time after applying the voltage and is responsible for the initial rise in $I(t)$. The current fluctuates as new connecting paths appear and disappear, ultimately falling towards 0 as the system dries up completely. This seems to be a plausible explanation why $I(t)$ falls monotonically in CP, but increases first, fluctuates and then falls for CN. Probably small fluctuations in the $N(t)$ vs. $t$ curves (figure(\ref{Ntsat})) for CN which are more prominent at higher voltages are also a result of this effect. For $V$ = 5 and 8 volts, less water issues at the center and $N(t)$ curves vary smoothly with time upto $N_{sat}$.

  The proposal given above is verified by performing a new set of experiments. We devise a set up which inverts the profile of the sample, making it lowest at the center and highest at the edge as follows.
  
   Two electrodes, one  in the form of a short cylinder (height $\sim 2.7 cm$, diameter $\sim 2 cm$) made of aluminium sheet and the other a naked aluminium wire of diameter $\sim 0.45 mm$ are connected to the DC power supply (figure(\ref{setup} B1, B2)). The cylinder is placed on an acrylic sheet and the aqueous solution of laponite is poured inside it. It is expected that the meniscus of the solution will be concave with a curvature approximately inverse to the droplet. In reality the curvature of the convex droplet is somewhat greater than that of this concave meniscus. Figure(\ref{setup} A and B) illustrates the two cases. If the argument we just outlined is correct, current variation in CP and CN configurations for the concave meniscus should be the reverse of what was seen for the convex droplet. That is exactly what we observe. 
   
   For the concave meniscus, current approaches zero from a maximum value for central electrode negative (CN) and shows a peak increasing from a smaller value to a maximum and then falling towards zero, for central electrode positive (CP). The nature of current fall through the sample for this case is almost opposite to that for the droplet, as shown in figure(\ref{current_droplet_cp_cn} b1, b2). So CP and CN behavior for the convex droplet (figure(\ref{current_droplet_cp_cn}) a1, a2) and the concave meniscus (figure(\ref{current_droplet_cp_cn}) b1, b2) are reverse of each other.
   
 We also checked the current variation with time for a large circular system,  where the sample surface is reasonably flat, for both CP and CN configurations. For this we used a poly propylene petri-dish of 10 cm diameter and the arrangement was same as in the previous study \cite{tajkera}. In this radial system, the field is non-uniform, being much stronger at the center compared to the periphery.
   
    In CP condition there is no water flow from outer to central electrode as the field near the outer electrode is very weak. Due to fast gelation near the central electrode, the circuit becomes effectively open and the current falls rapidly from a maximum towards zero.
    
     In CN condition, there is more and more water flow from central electrode towards the outer periphery. It can actually be seen that whenever a fresh stream of water from the center reaches the outer boundary a new peak is observed. The current fluctuates as small streams connect the two electrodes at different points and then dry out. The current finally falls to zero as the whole sample dries. So these experiments support our proposed explanation for the $I(t)$ variation in CP and CN configurations.
      
\subsection{Energy dissipated $E_{dis}$} 
While there is a current flowing, the power consumption primarily determines the evaporation rate and hence onset of crack formation. With $V$ held constant, the enhanced current in CN, would then account for faster crack formation i.e. lower $t_c$. We calculate the total energy dissipated ($E_{dis}$) in time $t_c$ as
 \begin{equation}
 E_{dis} = \int_0^{t_c} VI(t) \mathrm{d}t
 \end{equation}
 
 $E_{dis}$ is calculated numerically from the experimental data for $I(t)$ using the trapezoidal rule, for the different applied voltages and for both CP and CN configurations. We find that in all cases $E_{dis}$ has a nearly constant value. Only the point corresponding to 5V in the CP case deviates somewhat. However if $t_c$ for 5V CP is taken from the fitted curve in figure(\ref{tc_vs_appliedv}) shown as a \#, the corresponding $E_{dis}$ comes out much closer to the average which is $\sim 4.624 Joule$. The order of fluctuations is measured by standard deviation that takes the value 0.29. There might have been an error in identifying $t_c$ for 5V in CP.
 
 \subsection{Effect of varying $\tau$}
Even a very short duration of application of field (with $\tau$ of the order of seconds) leaves its signature on the pattern of cracks which appear much later. 
For a given $V$, as $\tau$ (which is of course less than $t_c$) decreases we have to wait longer for the cracks to appear, the time of appearance in this case is designated as $t_a$.
For both CP and CN conditions the patterns are affected when the duration of application of field ($\tau$) is as small as $\sim$ 15 sec, for all the voltages studied. The only exception being CN for 5V. For this case  $\tau$ needs to be at least $\sim 30 s$ to affect the crack pattern. For lower $\tau$, cracks appear randomly. For a fixed voltage the details of the final crack formation ($t_a$ and $n_f$) depend on $\tau$ as shown in figure(\ref{tau-variation-cp-cn}).
 
\begin{figure}[h]
\begin{center}
\includegraphics[width=16.0cm, angle=0]{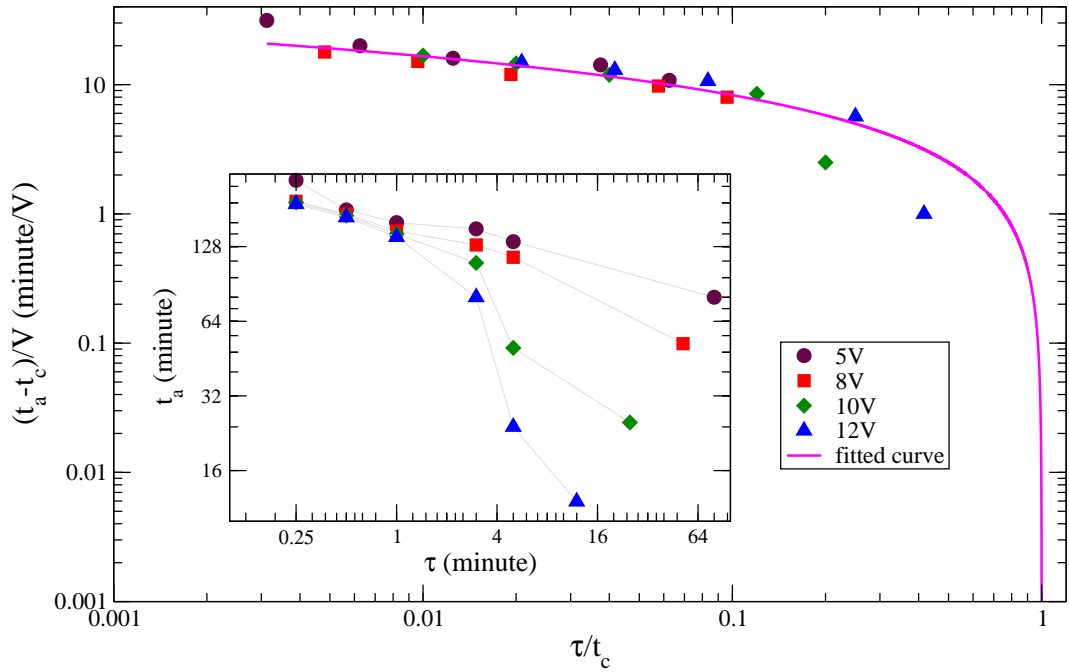}
\end{center}
\caption{The inset shows experimental $t_a$ vs. $\tau$ plots for different $V$ for CP. ($t_a$ - $t_c$)/$V$ vs. $\tau$/$t_c$ data for CP for different voltages ($V$) collapse to the single master curve represented by equation(\ref{ta}).}
\label{centre+ve_ta}
\end{figure}

\begin{figure}[h]
\begin{center}
\includegraphics[width=16.0cm, angle=0]{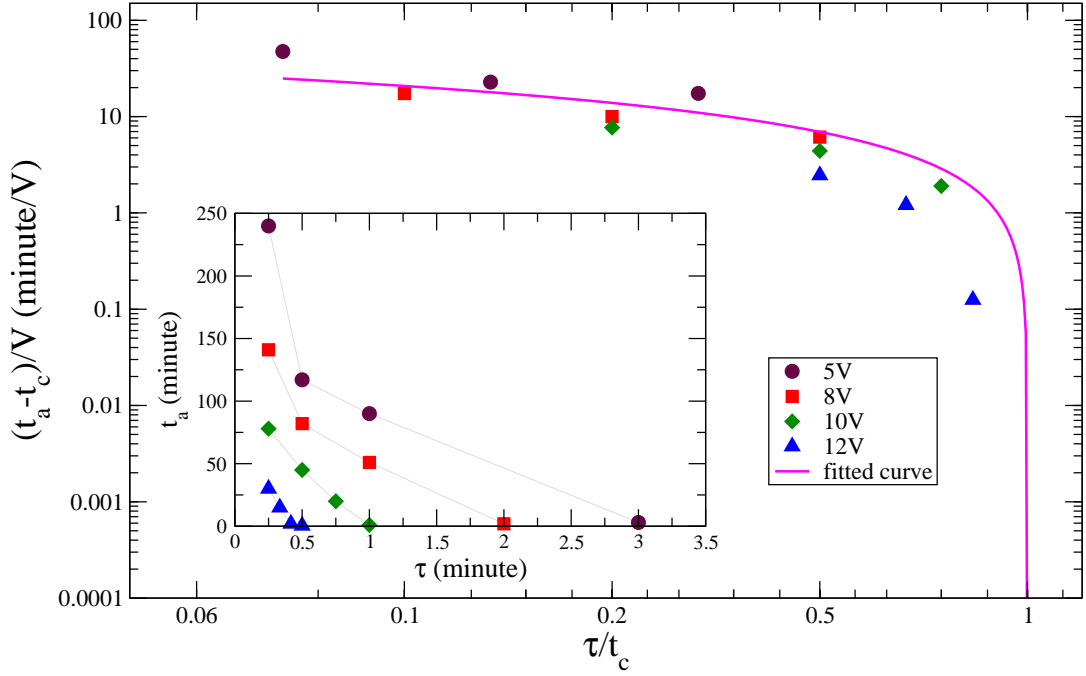}
\end{center}
\caption{The inset shows experimental $t_a$ vs. $\tau$ plots for different $V$ for CN. ($t_a$ - $t_c$)/$V$ vs. $\tau$/$t_c$ data for CN for different voltages ($V$) collapse to the single master curve represented by equation(\ref{ta}).}
\label{centre-ve_ta}
\end{figure}

 Crack patterns appear earlier as  $\tau$ is increased with $V$ constant, for both CP and CN conditions. Again for a fixed $\tau$, larger the magnitude of voltage, smaller is the time of appearance of first crack ($t_a$) for both CP and CN. For a fixed duration of application of field $\tau$ and for a fixed voltage ($V$), cracks appear faster for CN condition than for CP condition.  
 The insets in figures \ref{centre+ve_ta} and \ref{centre-ve_ta} show the plots of $t_a$ vs. $\tau$ for CP and CN  respectively. Here $t_a$ and $\tau$ are in minutes.

 \subsection{Variation in the final number of cracks $n_f$ with $V$ and $\tau$}
\begin{figure}[h]
\begin{center}
\includegraphics[width=16.0cm, angle=0]{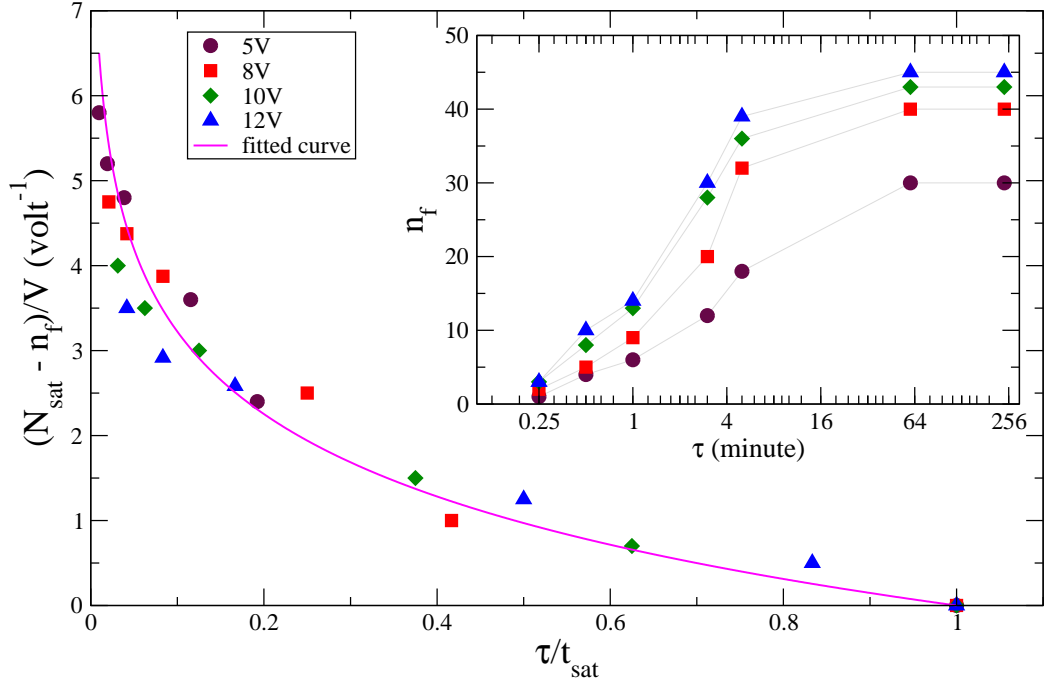}
\end{center}
\caption{The inset shows experimental $n_f$ vs. $\tau$ plots for different $V$ for CN. ($N_{sat}$ - $n_f$)/$V$ vs. $\tau$/$t_{sat}$ data for CN, for different voltages ($V$) collapse to the single master curve represented by equation(\ref{crnum}).}
\label{crack_num_cn_lin}
\end{figure}

 Interesting variation is displayed by the final number ($n_f$) of cracks as $\tau$ and $V$ are varied. For CP the cracks appear from the central electrode. In this set up, $n_f$ remains more or less constant ($\sim 3$ or very rarely 4), for all voltages ($V$) and for all $\tau$ as shown in figure(\ref{tau-variation-cp-cn} a1, a2). But a completely different picture is observed for CN condition. Here for a fixed $V$, $n_f$ increases with $\tau$ as shown for $V$ = 10 Volts in figure(\ref{tau-variation-cp-cn} b1, b2 and b3).  For the same $\tau$, $n_f$ increases with the magnitude of voltage ($V$). 
  The inset in figure(\ref{crack_num_cn_lin}) shows a plot of $n_f$ vs. $\tau$.

 As $\tau$ approaches $t_{sat}$, $n_f$ reaches the saturation value $N_{sat}$ for that particular $V$. The spacing between the cracks in this condition is
 \begin{equation}
 \lambda_{min}(V) = 2 \pi R_{out}/N_{sat}(V)
 \end{equation}
 $R_{out}$ being the radius of the outer electrode. We find $\lambda_{min}(12V) = 1.4 mm$.
 
 For the CP configuration where cracks appear from the central electrode, $n_f$ never exceeds 4. This is understandable since the central electrode being a thin wire of diameter $\sim 0.45 mm$, the perimeter $\sim 1.414 mm$ does not have enough space for more cracks. So the perimeter of the inner electrode is comparable to $\lambda_{min}(12V)$ and in this set up growth of more cracks from the central electrode is not possible.

\subsection{Scaling Relations} 

We show here that for CP and CN, $t_a$ with varying $V$ and $\tau$ can be scaled onto single master curves on appropriate transformations. For CN, $n_f$ with varying $V$ and $\tau$ can be similarly scaled onto a single master curve. 

The difference between $t_a$ and $t_c$ decreases as $\tau$ is increased. If we plot $(t_a - t_c)/V$ against $\tau/t_c$, the data for different $V$ and $\tau$ are found to collapse onto a single master curve, as shown in figure(\ref{centre+ve_ta}) for CP and in figure(\ref{centre-ve_ta}) for CN. The curves can be fitted by the simple empirical relation of the form

\begin{equation}
 \frac{t_{a}(\tau)-t_{c}}{V} = a - b.ln(\frac{\tau}{t_c})
  \label{ta}
 \end{equation}
 
 In both cases the boundary condition $\tau = t_c$ must lead to $a$ = 0, so the only free parameter to fit is $b$. We find
  $b$= 216 s/V for CP and $b$= 600 s/V for CN give best fits to the data. Here of course,  $\tau$ is always $\leq$ $t_c$, $t_c$ being the time at which first crack appears when the field is on throughout the experiment. 
  
  All data for the final number of cracks $n_f$ (for CN) scale to a similar master curve, when we plot $(N_{sat} - n_f)/V$ against $\tau/t_{sat}$ as shown in 
 figure(\ref{crack_num_cn_lin}). This curve follows the empirical rule
 
 \begin{equation}
 \frac{N_{sat}-n_f(\tau)}{V} = c - d.ln(\frac{\tau}{t_{sat}})
  \label{crnum}
 \end{equation} 
 
 The boundary condition leads to $c=0$, since for $\tau=t_{sat}$ the right hand side of the equation must be 0. 
Best fit to experimental data is obtained for $d$ = 1.4$V^{-1}$. Here  $\tau$ takes values $\leq$ $t_{sat}$. For CP $n_f$ is always around 3 for any $V$ and $\tau$, so this exercise is not required.

The scaling relations imply that
for any arbitrary $V$ and $\tau$, knowing $t_c$ from the equation(\ref{tc}) and $t_{sat}$ from figure(\ref{Ntsat}), we can predict the first time of appearance $t_a$ of a crack in CP or CN configuration as well as $n_f$ for CN by using the master curves. For CP, the final number of cracks is always 3 (very rarely 4), so $n_f$(CP) is known anyway.

 \section{Discussion}
      
 The most significant finding of our study is that several characteristic features related to the crack patterns in laponite droplets under a DC electric field obey the simple master equation, when the variables are appropriately scaled 
 \begin{equation}
 Y = - B. ln(X)
 \label{master}
 \end{equation}
 
 The 3 sets of variables which satisfy this equation are
 \begin{itemize}
 \item{$X = \frac{\tau}{t_c}$ and $Y = \frac{t_a - t_c}{V}$ for CP}
 \item{$X = \frac{\tau}{t_c}$ and $Y = \frac{t_a - t_c}{V}$ for CN}
 \item{$X = \frac{\tau}{t_{sat}}$ and $Y = \frac{N_{sat}-n_f}{V}$ for CN}
 
 \end{itemize}
 
 The first time of appearance of cracks $t_c$, when a given voltage $V$ is applied continuously, is itself a decreasing exponential function of the voltage V (equation(\ref{tc})). So the master equation(\ref{master}) can be used to predict $t_a$ for both CN and CP configurations, knowing $\tau$ and $V$. For CN, we can also predict $n_f$ from the master equation(\ref{master}), and of course as long as the central electrode is a thin wire, we know that for CP, $n_f$ is most likely to be 3. It would be interesting to see if these relations work for samples of different size and if so, how the empirical constants depend on system size. We expect a limitation to the system size as larger droplets will not have surfaces of the same shape.
 
The master curves for the first two cases are shown as double logarithmic plots (figures (\ref{centre+ve_ta}) and (\ref{centre-ve_ta})), since the time scales involved vary over orders of magnitude. In this format, the point corresponding to $\tau$ = $t_c$ (or $t_{sat}$ as the case may be), becomes undefined and cannot be shown on the graph. However, the master curve clearly tends towards $Y = 0$. The graph for $n_f$ is more conveniently displayed in linear scale figure(\ref{crack_num_cn_lin}), since the crack numbers do not vary over a very wide range. Here the point corresponding to the limiting value $X$ = 1 can plotted. Differentiating equation(\ref{master}) we may write

\begin{equation}
\frac{dY}{dX} = - \frac{B}{X}
\end{equation}

So, according to this law the rate of change of $Y$ with $X$ is inversely proportional to $X$ and the single fitting parameter $B$ can be interpreted as the magnitude of the negative slope of the master curve at the limiting value $X = 1$. It may also be interpreted as the constant value of $ (\frac{dY}{dX})X$ for $0 < X \leq 1$. For $X \rightarrow 0$, $Y \rightarrow \infty$, as it should (figure(\ref{crack_num_cn_lin})). 
In the linear plot (figure(\ref{crack_num_cn_lin})) the master curve shows a definite non zero negative slope at $X = 1$.
 
 Understanding the physical origin of the simple law given by equation(\ref{master}) is a major challenge. Obviously, the times $t_c$ and $t_{sat}$ represent important characteristic time scales for the desiccation and crack formation processes. The data collapse for different $V$, indicates that the difference $t_{diff} = (t_a - t_c)$, i.e. how long one has to wait after $t_c$ for cracks to appear after switching off the field, scaled by $V$ is a logarithmic function of the reduced time $\frac{\tau}{t_c}$. A similar relation holds for the final number of cracks according to figure(\ref{crack_num_cn_lin}), when time is scaled by $t_{sat}$ .
   
\section{Conclusions}

We have demonstrated that a laponite sample shows an unmistakable history dependence  towards electric field induced crack formation, reminiscent of the `Nakahara effect'\cite{naka} for mechanical perturbation. A very short exposure to the electric field induces a crack pattern characteristic of that field geometry, though producing less cracks and with a time lag which scales with the field strength. It is as if the system `remembers' what kind of field has been applied to it, though there were no cracks when the field was switched off. For example, if a CN field is applied for a few seconds and switched off, the final crack pattern shows peripheral cracks only, cracks never appear from the center.

It is instructive to compare the experimental results presented here with the earlier work on larger circular systems in a radial electric field \cite{tajkera}. Several observations are similar e.g. the emergence of radial cracks from the positive electrode and water emanating from the negative electrode. Absence of cracks near the negative electrode is also a feature common to both the droplet and the larger circular system. However, the appearance of cross-radial cracks for the CN configuration\cite{tajkera} is not seen here. The droplet may be too small for significant field gradients to develop. A preliminary study with a stronger field gradient does show cross-radial cracks, further work is in progress. The quantification of different observations and relating them through empirical relations has been possible in the present work because the time taken for the whole drying process is at most 4-5 hours for the droplet. In case of large samples this time was 3-4 days, so environmental effects other than the imposed electric field have time to act, making it difficult to isolate the field effect and frame definite rules for it. The typical convex droplet shape adds a further dimension to the problem.

      There are some possible sources of error in the results reported, arising form the extreme simplicity of the set up. The circular peripheral electrode is shaped by hand, so it is only approximately circular. Moreover, each data point has to be determined from a new set of electrodes, because after passing current the electrodes corrode and cannot be reused. In addition there may be slight fluctuations in ambient temperature and humidity. It is quite amazing that in spite of these limitations we obtain rather simple and convincing results. Further experiments and theoretical studies are in progress to understand and establish these empirical rules on a more firm footing, also to investigate the effect of changing the geometry and size of the set up.  
            
 \section{Acknowledgement}
 TK thanks CSIR for providing a research grant. This work is supported by a joint Indo-Japan collaboration supported by DST-JSPS. Authors are extremely grateful to the Japanese collaborators - Akio Nakahara, So Kitsunezaki, Chiyori Urabe, Ooshida Takeshi, Takahiro Hatano, and Nobuyasu Ito, members of the Japanese team for visiting India and providing fruitful suggestions during extensive discussions, in particular for pointing out the significance of the current variation during the experiment. Authors thank T.R. Middya, Department of Physics, Jadavpur University, for support and useful discussion. Special thanks to Rockwood Additives for gifting the sample of Laponite RD.


\begin{thebibliography}{99}
\bibitem{review}Routh, A.F. Drying of thin colloidal films. {\it Reports on Progress in Physics} $\bf2013$, {\it 76}, 046603 (30pp).\\
\bibitem{buckling}Tsapis, N.; Dufresne, E. R.; Sinha, S. S.; Riera, C. S.; Hutchinson, J. W.; Mahadevan, L.; Weitz, D. A. Onset of Buckling in Drying Droplets of Colloidal Suspensions. {\it Physical Review Letters} $\bf2005$, {\it 94}, 018302.\\
\bibitem{moutushi}Choudhury, M.D.; Dutta, T.; Tarafdar, S. Pattern formation in droplets of starch gels containing NaCl dried on different surfaces. {\it Colloids and Surfaces A: Physicochemical and Engineering Aspects} $\bf2013$, {\it 432}, 110–118.\\
\bibitem{anarelli1}Annarelli, C.C.; Fornazero, J.; Bert, J.; Colombani, j. Crack patterns in drying protein solution drops.  {\it European Physical Journal E} $\bf2001$, {\it 5}, 599–603.\\
\bibitem{sobac} Sobac, B.; Brutin, D. Structural and evaporative evolutions in desiccating sessile drops of blood. {\it Physical Review E} $\bf2011$, {\it 84}, 011603.\\
\bibitem{brutin} Brutin, D.; Sobac, B.; Loquet, B.; Sampol, J. Pattern formation in drying drops of blood. {\it Journal of Fluid Mechanics} $\bf2011$, {\it 667}, 85-95.\\
\bibitem{brutin-langmuir}Carle, F.; Brutin, D. How surface functional groups influence fracturation
in nanofluid droplet dry-outs. {\it Langmuir} $\bf2013$, in Press.\\
\bibitem{evaporation}Tarasevich, Y.Y. Simple analytical model of capillary flow in an evaporating sessile drops. {\it Physical Review E} $\bf2005$, {\it 71}, 027301.\\
\bibitem{colsua14}Vancauenberghe, V.; Di Marco, P.; Brutin, D. Wetting and evaporation of a sessile drop under an external electric field: a review. {\it Colloids and Surfaces A: Physicochemical and Engineering Aspects} $\bf2013$, {\it 432}, 50–56.\\ 
\bibitem{lucas}Goehring, L.; Clegg, W. J.; Routh, A. F. Solidification and Ordering during Directional Drying of a Colloidal Dispersion. {\it Langmuir} $\bf2010$, {\it 26(12)}, 9269–9275. \\
\bibitem{daniels}Daniels, K.E.; Mukhopadhyay, S.; Houseworth, P.J.; Behringer, R.P. Instabilities in Droplets Spreading on Gels. {\it Physical Review Letters} $\bf2007$, {\it 99}, 124501.\\
\bibitem{pauchard1}Pauchard, L.; Parisse, F.; Allain, C. Influence of salt content on crack patterns formed through colloidal suspension desiccation. {\it Physical Review E} $\bf1999$, {\it 59(3)}, 3737–3740.\\
\bibitem{so}Kitsunezaki, S. Crack Growth and Plastic Relaxation in a Drying Paste Layer. {\it Journal of the Physical Society of Japan} $\bf2010$, {\it 79}, 124802.\\
\bibitem{lopes}Lopes, M. C.; Bonaccurso, E.; Gambaryan-Roismana, T.; Stephan, P. Influence of the substrate thermal properties on sessile droplet evaporation: Effect of transient heat transport. {\it Colloids and Surfaces A:Physicochemical and Engineering Aspects} $\bf2013$, {\it 432}, 64–70.\\
\bibitem{naka}Nakahara, A.; Matsuo, Y. Transition in the pattern of cracks resulting from memory effects in paste. {\it Physical Review E} $\bf2006$, {\it 74}, 045102(R).\\
\bibitem{pauchard}Pauchard, L.; Elias, F.; Boltenhagen, P.; Cebers, A.; Bacri, J.C. When a crack is oriented by a magnetic field. {\it Physical Review E} $\bf2008$, {\it 77}, 021402.\\
\bibitem{tajkera}Khatun, T.; Choudhury, M.D.; Dutta, T.; Tarafdar, S. Electric-field-induced crack patterns: Experiments and simulation. {\it Physical Review E} $\bf2012$, {\it 86}, 016114.\\
\bibitem{ac}Khatun, T.; Dutta, T.; Tarafdar, S. Crack Formation in Laponite Gel under AC Fields. arXiv:1304.5523[cond-mat.soft].\\
\bibitem{mal1}Mal, D.; Sinha, S.; Middya, T.R.; Tarafdar, S. Field induced radial crack patterns in drying laponite gel. {\it Physica A} $\bf2007$, {\it 384}, 182-186.\\
\bibitem{mal2}Mal, D.; Sinha, S.; Middya, T.R.; Tarafdar, S. Desiccation crack patterns in drying laponite gel formed in an electrostatic field. Applied Clay Science $\bf2008$, 39 , 106–111.\\ 
\bibitem{yakhno}Yakhno, T. Salt-induced protein phase transitions in drying drops. {\it Journal of Colloid and Interface Science} $\bf2008$, {\it 318}, 225-230.\\
\bibitem{bardakov}Bardakov, R.N.; Chashechkin, Yu.D.; Shabalin, V.V. Hydrodynamics of a Drying Multicomponent Liquid Droplet. {\it Fluid Dynamics} $\bf2010$, {\it 45(5)}, 141-155.\\
\bibitem{tarasevich}Tarasevich, Yu.Yu.; Pravoslavnova, D.M. Segregation in desiccation sessile drops of biological fluids. {\it European Physical Journal E} $\bf2007$, {\it 22}, 311-314.\\
\bibitem{poster}Khatun, T.; Dutta, T.; Tarafdar, S.; Nakahara, A.; Kitsunezaki, S.; Urabe, C.; Takeshi, O.; Ito, N. Crack formation in droplets and rectangular samples of laponite gel
dried under an electric field. Poster presentation at DROPLET-2013, 1st International workshop on wetting and evaporation, Marseilles, June 17-20, 2013.\\ 
\bibitem{net}Website: http://www.laponite.com/faqs.asp\\

 \end{thebibliography}
\end{document}